\documentclass[journal=inoraj,manuscript=article]{achemso}
\setkeys{acs}{articletitle = true}
\usepackage[version=3]{mhchem} 
\usepackage{chemformula}
\usepackage{amsmath}

\title{Hydrothermal synthesis of ordered corkite \ch{PbFe3(PO4)(SO4)(OH)6}, a S~=~5/2 kagomé antiferromagnet}
\author{Austin M. Ferrenti}
\affiliation{Department of Chemistry, The Johns Hopkins University, Baltimore, Maryland 21218, USA}
\alsoaffiliation{Institute for Quantum Matter, William H. Miller III Department of Physics and Astronomy, The Johns Hopkins University, Baltimore, Maryland 21218, USA}
\email{aferren2@jhu.edu}
\author{Vanessa Meschke}
\affiliation{Department of Physics, Colorado School of
Mines, Golden, Colorado 80401, USA}
\author{Shreenanda Ghosh}
\affiliation{Institute for Quantum Matter, William H. Miller III Department of Physics and Astronomy, The Johns Hopkins University, Baltimore, Maryland 21218, USA}
\author{Jackson Davis}
\affiliation{Institute for Quantum Matter, William H. Miller III Department of Physics and Astronomy, The Johns Hopkins University, Baltimore, Maryland 21218, USA}
\author{Natalia Drichko}
\affiliation{Institute for Quantum Matter, William H. Miller III Department of Physics and Astronomy, The Johns Hopkins University, Baltimore, Maryland 21218, USA}
\author{Eric. S. Toberer}
\affiliation{Department of Physics, Colorado School of
Mines, Golden, Colorado 80401, USA}
\author{Tyrel M. McQueen}
\affiliation{Department of Chemistry, The Johns Hopkins University, Baltimore, Maryland 21218, USA}
\alsoaffiliation{Institute for Quantum Matter, William H. Miller III Department of Physics and Astronomy, The Johns Hopkins University, Baltimore, Maryland 21218, USA}
\altaffiliation{Department of Materials Science and Engineering, The Johns Hopkins University, Baltimore, Maryland 21218, USA}
\email{mcqueen@jhu.edu}

\begin{document}

\begin{abstract}

Corkite, \ch{PbFe3(PO4)(SO4)(OH)6}, an understudied relative of the jarosite family of Heisenberg antiferromagnets, has been synthesized and its magnetic properties characterized for the first time. Relative to natural samples, synthetic corkite displays signatures in both infrared and Raman spectra of a more ordered arrangement of polyanion groups about the kagomé sublattice that retains inversion symmetry. Magnetic susceptibility measurements reveal that dried corkite undergoes a transition to a long-range, antiferromagnetically-ordered state below $T_N$~=~48 K, lower than that observed in the majority of jarosite phases, and indicative of further spin frustration. Curie-Weiss fitting of the measured magnetic susceptibility yields an effective magnetic moment of p$_{eff}$~=~6.29(1)~$\mu_B$/Fe$^{3+}$ and  $\theta_{CW}$~=~-526.0(1.1)~K, analogous to that observed in similar high-spin Fe$^{3+}$ systems, and indicative of strong antiferromagnetic coupling. Estimation of the change in magnetic entropy as a function of temperature from T~=~0~K to T~=~195~K, $\Delta S_{mag}$~=~14.86~$J\cdot mol_{Fe^{3+}}^{-1}K^{-1}$, is also in good agreement with the $\Delta S_{mag}$~=~Rln(2S+1)~=~14.9~$J\cdot mol^{-1}K^{-1}$ expected for a S~=~5/2 system. In comparison to the pure jarosites, where both structure and magnetism remain largely invariant upon a variety of chemical substitutions, the replacement of one sulfate group per formula unit with a higher-valent phosphate group applies additional steric and electronic pressure on the kagomé lattice in corkite, further frustrating the magnetic ground state of the material. Corkite thus represents both an outlier in the known body of jarosite-type materials, and an illustration of how existing structures may be further strained in the development of highly frustrated magnetic systems.

\end{abstract}

\section{Introduction}

Materials containing the kagomé lattice, consisting of a two-dimensional network of corner-sharing triangles, have long been studied for the novel phenomena they are often host to, including strongly correlated topological states\cite{zhang2021recent}, superconductivity,\cite{neupert2022charge} and most notably high degrees of magnetic frustration. \cite{shores2005structurally, hiroi2001spin, Meschke2021} One such family of materials, the jarosites, (\ch{AB3(XO4)2(OH)6}, where A = a monovalent or divalent cation, B = a trivalent cation, and \ch{XO4} = a divalent polyanion group) have long been studied for the impact of kagomé-sublattice vacancies on their magnetic properties \cite{Grohol2003,Basciano2007}. The majority of the jarosite phases possess metal-ligand-metal (M-L-M) bridging angles on the order of 130-140°, are known to order antiferromagnetically between \emph{T}~=~55-65~K, and are considered near-ideal Heisenberg antiferromagnets, with weakly ferromagnetic intralayer coupling and antiferromagnetic interplanar coupling \cite{Inami2000,Wills2000}. The strongly anisotropic magnetism in these materials arises from the Dzyaloshinskii-Moriya (DM) interaction, whereby the tilting of the Fe octahedra which comprise the kagomé lattice produces a canting of the spins about each triangular hole into a spin umbrella configuration \cite{yildirim2006magnetic}. Despite the high degree of variability in both the cation and capping groups which have been successfully substituted into the jarosite structure, the tilt angle and degree of distortion of the kagomé-layer \ch{FeO6} octahedra remains relatively consistent across the family. 

\begin{figure}
  \includegraphics[width=1.00\textwidth]{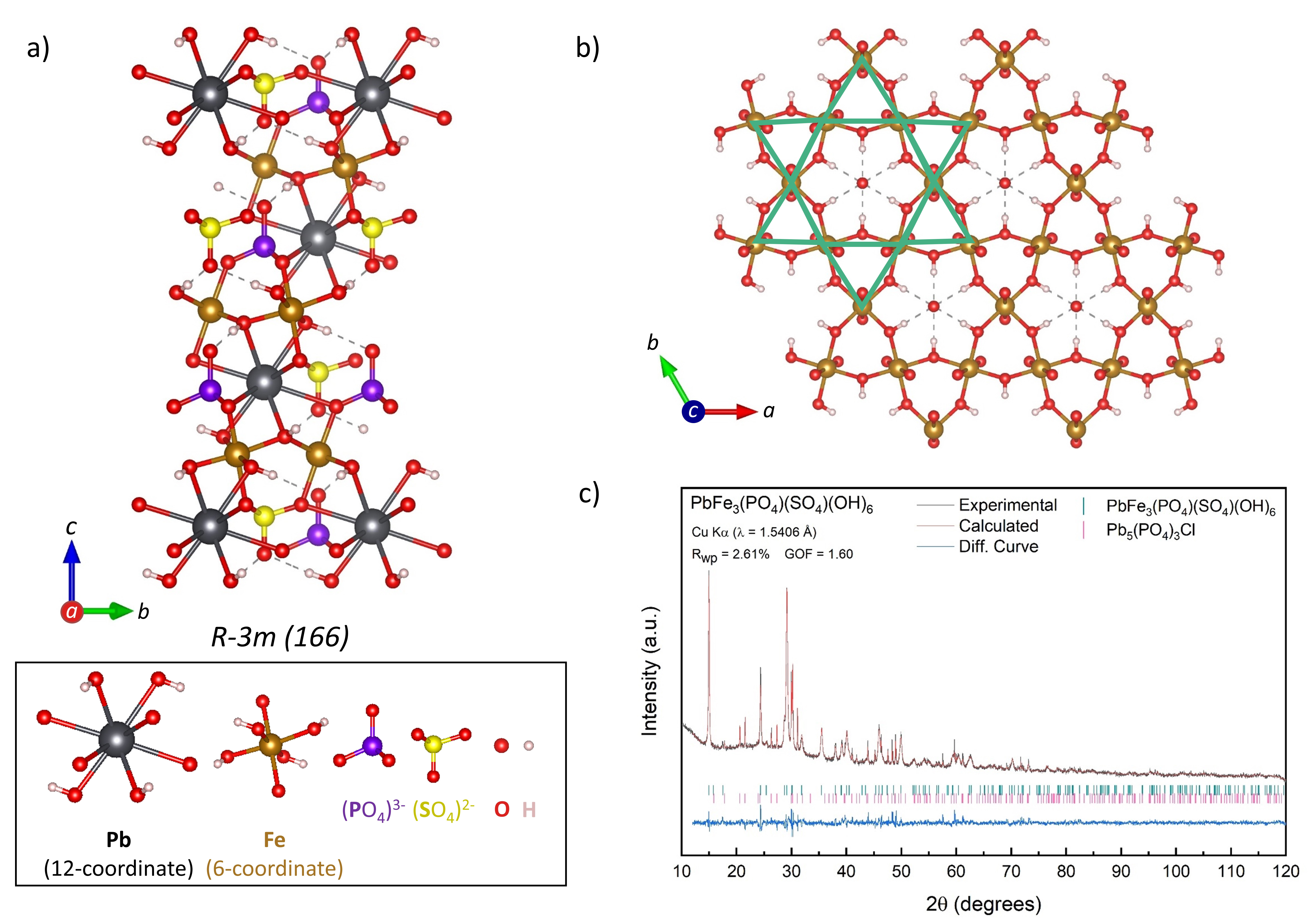}
  \caption{Crystal structure of corkite projected a) along the \emph{a} axis and b) along the \emph{c} axis. Each lead (black) and iron (gold) atom is 12-fold and 6-fold coordinate, respectively, with 2/3 hydroxide and 1/3 oxygen (red) bridging ligands. The kagomé plane (emphasized in green) is bound by both highly-coordinated lead atoms and alternating phosphate (purple) and sulfate (yellow) capping groups. c) pXRD pattern of a representative synthetic corkite sample (black). Two-phase simulated Rietveld refinement profile for \ch{PbFe3(PO4)(SO4)(OH)6} (space group \emph{R$\Bar{3}$mH}) and \ch{Pb5(PO4)3Cl} (space group \emph{P63/m}) (red). Difference curve between the experimental and simulated patterns (blue). Refinement of the experimental pattern indicates that the best samples possess approximately 91.0\% of the corkite phase by weight.}
  \label{XRD}
\end{figure}

However, several jarosite phases have been reported which contradict this observed trend. The hydronium and deuterated jarosite analogues, owing to the favorability of interactions between \ch{H3O+} and \ch{OH-} ions in the structure, instead exhibit a spin glass transition around $T_g$~=~15~K \cite{And2007}.
In order to maintain charge neutrality, jarosite phases containing divalent cations have only a 1/2 A-site occupancy, but otherwise possess quite similar structural and magnetic properties to those possessing full A-site occupancy. The Pb-analogue plumbojarosite, \ch{Pb_{0.5}Fe3(SO4)(OH)6}, is observed to order antiferromagnetically below $T_N$~=~56.4~K, which is slightly lower than other members of the family, and likely due to increased steric effects from the larger \ch{Pb^{2+}} cation.\cite{And2005}.  Beudantite, \ch{PbFe3(SO4)(AsO4)(OH)6}, containing alternating sulfate and larger arsenate groups, has only been characterized from a heavily \ch{H3O+}-substituted natural sample, and was observed to order at around $T_N$~=~45~K, indicating that the additional steric pressure on the kagomé plane further frustrates the spins, thereby suppressing long-range order \cite{wills2004interlayer}. A recent computational study suggested that corkite, \ch{PbFe3(PO4)(SO4)(OH)6}, a related mineral with a similar calculated ground state Ising spin configuration to the jarosites, would benefit from magnetic characterization to determine whether it behaves as a Heisenberg antiferromagnet, or if it exhibits more interesting magnetism at low temperatures \cite{Meschke2021}. The structure of corkite has been previously probed in both natural and synthetic samples (Figure \ref{XRD}a, \ref{XRD}b), however its magnetic properties have not yet been reported.\cite{Giuseppetti1987corkite, frost2011vibrational} While pure phase specimens of corkite would not contain the rather reactive \ch{H3O+} species, it is not known whether the substitution of one sulfate group per formula unit with a higher-valent, larger phosphate group could disrupt the interlayer couplings observed in the pure jarosites,  producing a more ideal Heisenberg antiferromagnet.

Here we report the hydrothermal synthesis of synthetic corkite, as well as the characterization of its structural and magnetic properties. A combination of powder x-ray diffraction and IR and Raman spectroscopy measurements are used to confirm prior structural assignments of both natural and synthetic corkite samples in the \emph{R$\Bar{3}$m} space group (166). Magnetic susceptibility and heat capacity measurements are also performed and show synthetic corkite to order antiferromagnetically below T$_N$~=~48~K, with an effective magnetic moment on the order expected for an Fe$^{3+}$, S~=~5/2 system. Our results show corkite to be a frustrated Heisenberg antiferromagnet, and provide a deeper understanding of how the split polyanion site occupancy, and subsequently larger steric and electronic strain, impacts the magnetic behavior of corkite, relative to the broader family of jarosite materials.

\section{Results and discussion}
\subsection{Synthesis}
Corkite samples were synthesized in Teflon-lined hydrothermal vessels at elevated temperatures. Heating and cooling ramp rates for all syntheses were set at 100°C per hour. The precursor \ch{Pb5(PO4)3Cl} was synthesized in two steps. First, stoichiometric quantities of PbO (Alfa Aesar, $\sim$99.9\%) and \ch{NH4H2PO4} (VWR, ACS Grade) were thoroughly ground and heated in an uncovered platinum crucible for 2 days at 1000°C. Platinum was chosen due to its lower reactivity with PbO, compared to alumina. The product \ch{Pb3(PO4)2} powder was then ground in a 3:2 molar ratio with \ch{PbCl2} (Alfa Aesar, 98\%) and heated for 2 days at 850°C in a covered alumina crucible. \ch{FePO4} was also synthesized in two stages. First, equimolar amounts of \ch{NH4H2PO4} and \ch{Fe(NO3)3} $\cdot$ \ch{9 H2O} (Aqua Solutions, ACS Grade) were dissolved in deionized water, stirred until clear, then boiled until dry. The resulting powder was dried for 24 hours at 400°C, and then for an additional 24 hours at 700°C. Corkite samples were produced by the combination of a ground mixture of 0.95 mol \ch{Pb5(PO4)3Cl}, 2.00 mol \ch{FePO4}, and 1.67 mol \ch{Fe2(SO4)3} $\cdot$ \ch{9 H2O} (Aqua Solutions, Reagent Grade) with 9.67 mol \ch{FeCl3} $\cdot$ \ch{6 H2O} (Alfa Aesar, 98\%). All powders were added to the Teflon liner, filled to $\sim$70\% with distilled \ch{H2O}, further acidified with 3 drops of $\approx$5M \ch{H2SO4} solution, sealed, and heated for 4 days at 98°C. 

Samples produced by the above method consistently resulted in large \ch{Pb5(PO4)3Cl} and \ch{PbSO4} impurities, the latter of which could be removed by stirring the product mixture in a concentrated 12.2M sodium acetate solution in 6 minute increments. Due to the higher stability of \ch{Pb5(PO4)3Cl} in acidic solution relative to corkite, this impurity could only be partially removed with a 1.0 M KOH solution without decomposition of the target phase and formation of amorphous iron oxide/hydroxide species. As \ch{Pb5(PO4)3Cl} was present in the system both pre- and post-reaction, several syntheses were attempted with reduced precursor concentrations. However, all resulted in both a reduction in corkite yield and the formation of additional impurity phases. A 5\% deficiency of \ch{Pb5(PO4)3Cl} was ultimately found to produce the purest product mixture ($\sim$91\% corkite). For this reason, and owing to the non-magnetic nature of \ch{Pb5(PO4)3Cl}, all characterization measurements have been performed on representative samples containing both phases.

Reaction times shorter than 4 days typically yielded only minimal quantities of the target corkite phase, instead consisting largely of \ch{Pb5(PO4)3Cl} and \ch{PbSO4}. Here, the lower solubility of \ch{Pb5(PO4)3Cl} limits the incorporation of Pb$^{2+}$ ions into the corkite structure and thus also its formation, but also prevents the immediate precipitation of large quantities of \ch{PbSO4}. Reaction times longer than 4 days ultimately resulted in less corkite in the product mixture and no \ch{Pb5(PO4)3Cl} precipitation, as well as the formation of various lead and iron phosphate impurity phases. When the precursor \ch{Pb5(PO4)3Cl} has fully dissociated, \ch{PbSO4} precipitation becomes more favorable relative to corkite, leaving fewer sulfate groups to stabilize the corkite structure, and ultimately producing only a S-deficient, more-heavily disordered analogue. The formation of corkite is therefore most favorable only for reaction times close to 4 days, where the relative solubilities of \ch{Pb5(PO4)3Cl}, \ch{PbSO4}, and corkite are more balanced, and sufficient concentrations of all constituent ions are present in solution.

In order to better approximate the non-magnetic contribution to the measured heat capacity, the Ga-analogue of corkite (\ch{PbGa3(PO4)(SO4)(OH)6}) was synthesized under conditions identical to those described for the parent compound. Precursor \ch{GaPO4} was synthesized by heating a well-ground, stoichiometric mixture of \ch{Ga2O3} (NOAH, 99.995\%) and \ch{NH4H2PO4} (VWR, ACS Grade) in air for 24 hours at 800°C. A ground mixture of 3 mol \ch{Pb5(PO4)3Cl}, 6 mol \ch{GaPO4}, 5 mol \ch{Ga2(SO4)3} $\cdot$ \ch{H2O} (Thermo Scientific, 99.999\% (metals basis)), and 14.5 mol \ch{Ga2O3} (NOAH, 99.995\%) was added to a Teflon-lined hydrothermal vessel, sealed, and heated for 4 days at 98°C. The resulting product was stirred for 6 minutes in a concentrated 12.2M sodium acetate solution to remove the \ch{PbSO4} impurity phase. Relative to the majority of single-step precipitation Fe-corkite synthesis trials, the \ch{Pb5(PO4)3Cl} impurity in all Ga-corkite trials was found to be significantly reduced without a secondary washing step, comprising only 8-9\% of the sample by weight. As a result, the product mixture was only stirred in 12.2M sodium acetate solution once for 6 minutes. 
Unlike the majority of reported jarosite systems, Al-corkite could not be synthesized in any quantifiable amount, at least under the chosen reaction conditions.

Synthetic jarosites initially produced via the traditional single-step precipitation method were frequently found to yield samples with up to a 15\% Fe deficiency on the kagomé sublattice, as well partial A-site substitutions by \ch{H3O+}.\cite{Wills2000} Later methods employing a slower, two-step precipitation method ultimately produced synthetic jarosites with full B-site occupancies and negligible A-site substitutions.\cite{bartlett2005long} Due to the strong dependence of the magnetic properties of other jarosite-family phases on the occupancy of both the A- and B- sites, the latter synthesis route was also explored in an attempt to discourage substitution by \ch{H3O+}. When Fe metal was instead used as the sole source of magnetic cations, the speed at which Fe$^{3+}$ entered the solution was reduced and the amount of corkite produced decreased significantly relative to trials employing more soluble species. This, as well as the observed time-dependence of corkite formation via the single-step precipitation method, suggests that corkite is the metastable product in this reaction system, in contrast with the more thermodynamically-stable jarosites produced by the same method.

\subsection{Characterization Methods}
Powder x-ray diffraction (pXRD) patterns were collected on a laboratory Bruker D8 Focus diffractometer with LynxEye detector and Cu K$\alpha$ radiation in the 2$\theta$ range from 5-120 degrees. Rietveld refinements on pXRD data were performed using Topas 5.0 (Bruker). Structures were visualized with the Vesta 3 program. \cite{momma2011vesta}

Magnetization data was collected on a Quantum Design Magnetic Property Measurement System (MPMS3). Magnetic susceptibility was approximated as magnetization divided by the applied magnetic field ($\chi\approx M/H$). Representative dry magnetization data was collected from synthetic corkite samples dried overnight at 150°C in air. Magnetization data for K,S-jarosite, K,Se-jarosite, and Pb-jarosite was extracted from previous studies (Ref. 9 and Ref. 12), and $\chi_0$ values were chosen such that each exhibited ideal Curie-Weiss behavior at high temperatures [K,S-jarosite: $\chi_0$~=~0.0007, K,Se-jarosite: $\chi_0$~=~-0.0060, Pb-jarosite: $\chi_0$~=~-0.0092]. The magnetization of both K,Se-jarosite and Pb-jarosite was also directly scaled by a factor of 0.44 and 0.37 respectively to allow for direct comparison. 

Heat-capacity data was collected on a Quantum Design Physical Property Measurement System (PPMS) using the semi-adiabatic pulse technique with a 1\% temperature rise and over three time constants. The change in magnetic entropy as a function of temperature, $\Delta S_{mag}$, was approximated as $\Delta S_{mag}$~=~$\int C_{p}/T\,dT$ of the measured $C_{p,mag}$ for corkite ($C_{p, total, Fe-corkite}$ - $C_{p, total, Ga-corkite}$), from \emph{T}~=~0--212 K. The entropy rise from T~=~0~K to 2~K was estimated from linear extrapolation over this range.

Simultaneous thermogravimetric analysis/differential thermal analysis (TGA/DTA) was performed using a TA Instruments Q600 SDT. The samples were loaded into pre-dried alumina pans and ramped quickly to 50°C. After holding for several minutes, the system was then ramped to 600°C at a rate of 10°C/minute under inert \ch{N2} flow (100 mL/minute). The sample was then allowed to cool to room-temperature under \ch{N2} gas flowing at the same rate.

Fourier-transform infrared (FTIR) measurements of synthetic corkite powder were conducted using a Thermo Nicolet Nexus 870 ESP FTIR with a Golden Gate KRS5 ATR accessory from 500 to 4000 cm$^{-1}$. Micro-Raman spectra of synthetic corkite powder were collected using a Horiba JY T64000 spectrometer equipped with an Olympus microscope in single monochromator mode under a 50x objective magnification. The samples were excited using the 514.5 nm line of a Coherent Innova 70C Spectrum laser, and the spectra collected using a Horiba Symphony II CCD detector. No signs of laser-induced degradation of the sample were observed over the course of each measurement. Representative dry sample spectra were collected from corkite samples dried overnight at 200°C in air.

\subsection{Structural characterization}
To determine the phase purity of the synthesized polycrystalline corkite samples, pXRD measurements and subsequent Rietveld refinements were performed. The pXRD pattern of a synthetic corkite sample of representative purity (Figure \ref{XRD}c) shows the presence of two distinct phases, as indicated by the contrasting peak widths observed. Broader reflections could be attributed to the target corkite phase crystallizing in the trigonal \emph{R$\Bar{3}$m} space group (166), while the sharper reflections were attributed to a hexagonal \ch{Pb5(PO4)3Cl} impurity. In several samples, a weak reflection also appeared at around $2\theta$~=~8.8 degrees, which seemed to correspond to a complex Pb-/Fe-phosphate phase and could not be removed via washing in either strongly acidic or basic media. The two-phase refinement of the purest corkite samples indicates that the non-magnetic \ch{Pb5(PO4)3Cl} impurity is present at 9.5(1)\% by weight (Tables \ref{Rietveld_table1}-\ref{Rietveld_table2}). The refined corkite lattice parameters obtained were a~=~b~=~7.3078(3)~\AA, c~=~16.855(2)~\AA, $\alpha$~=~$\beta$~=~90°, $\gamma$~=~120°. 
Slight deficiencies observed in refined site occupancies, as well as relatively large thermal displacement parameters, can likely be attributed to slight disorder in the material, as has been commonly observed in other synthetic members of the jarosite family.\cite{Grohol2003, Basciano2007} Drying of the samples produced no clear change in the measured diffraction pattern, indicating that no significant structural changes occur upon heating to 200°C.

Significant disagreement has thus far accompanied determination of the true structure of corkite, with the majority of recent studies on natural samples concluding that the centrosymmetric \emph{R$\Bar{3}$m} space group (166) is a better fit than the noncentrosymmetric \emph{R3m} group (160). \cite{kolitsch1999hinsdalite, taylor1997crystal, szymanski1988crystal} However, the elemental composition of natural corkite samples typically differs from the nominal stoichiometry, often possessing significant disorder on the phosphorus and sulfur sites. The sole report of structural studies on a sulfate-rich synthetic sample of the material only notes that the \emph{R3m} space group would be appropriate if the phosphate and sulfate groups occupy alternate sites in an ordered manner.\cite{baker1962mineral} As the two space groups differ only by the absence and presence of a net electric dipole moment, respectively, pXRD is not a viable means of distinguishing between them. 

Refinement of the experimental diffraction pattern in both the \emph{R$\Bar{3}$m} and \emph{R3m} structure types resulted in visibly similar fits to the experimental data. The refinement shown in Figure \ref{XRD} was carried out in the nonpolar \emph{R$\Bar{3}$m} space group ($R_{wp}$ value 2.61\%), as would be expected for synthetic corkite, with ordered alternating phosphate and sulfate groups about the kagomé plane. However, refinement in the polar \emph{R3m} space group ($R_{wp}$ value 2.51\%) also resulted in analogous calculated site occupancies and thermal displacement parameters. The conflicting structural assignments made in the literature can be interpreted as arising from varying degrees of disorder on the \ch{[PO4]^{3-}} and \ch{[SO4]^{2-}} sites bounding each kagomé layer, producing highly variable distributions of disordered and ordered regions which could average to produce a net dipole of variable magnitude and a subsequently more polar structure.

\begin{figure}
  \includegraphics[width=1.00\textwidth]{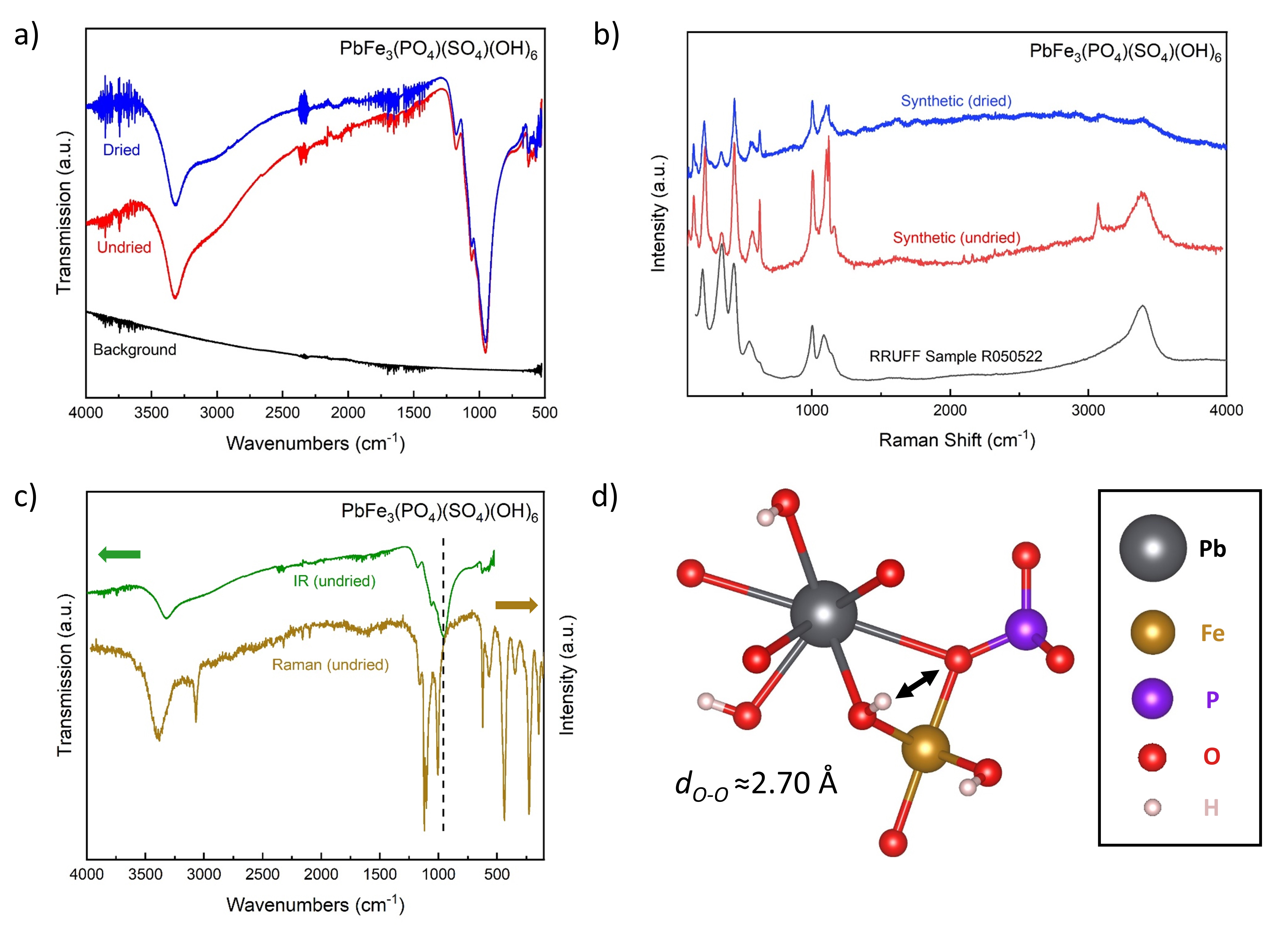}
  \caption{a) Infrared spectra of representative undried (red) and dried (blue) synthetic corkite powder samples. The collected background is shown in black for comparison. b) Raman spectra of undried (red) and dried (blue) synthetic corkite samples, as well as a representative natural corkite sample (RRUFF{\texttrademark} Project, Sample 050522 - black). c) Comparison of the infrared (green) and Raman (gold) spectra collected on a representative undried synthetic corkite powder sample. The black dotted line highlights the most intense mode in the IR spectrum, which has no corresponding mode in the Raman spectrum of the same sample. No normal vibrational modes are observed to overlap, indicative of a structure with inversion symmetry. d) OH$\mathord{\cdot\cdot\cdot}$O hydrogen bond attributed to the peak observed at 3069~cm$^{-1}$ in the undried synthetic Raman spectrum.}
  \label{Spec_Data}
\end{figure}

In order to draw further comparisons to previous spectroscopic studies of natural corkite, infrared and Raman spectra were collected from representative synthetic corkite samples, both before and after drying. The majority of features in the measured infrared spectra, shown in Figure \ref{Spec_Data}a and summarized in Table \ref{IR_data}, are significantly sharper than those observed for the natural specimens, which is indicative of a more ordered structure and subsequently fewer distinct coordination environments. This is also evident in the OH-stretching region (here from 2800-3500 cm$^{-1}$), where two broad humps are observed. In natural samples only one broader feature appears, due to the wider distribution of OH-stretching frequencies which arise from the higher degree of disorder. The majority of band assignments were made in agreement with those made by Frost \emph{et al}\cite{frost2011vibrational} based on the symmetry of the ordered structure. Although drying of the synthetic sample would be expected to result in the disappearance of the water stretching band at $\sim$1630 cm$^{-1}$, the feature is present in both spectra, possibly due to either the incomplete removal of water from the sample upon drying, or to minor hydration of the sample over the course of the measurement. This band is common among other jarosite materials, both natural and synthetic. However as the ordered corkite structure does not possess a monovalent cation which can be as easily substituted, significant water incorporation is less likely \cite{grohol2002hydrothermal,bishop2005visible}. A background measurement performed without a sample present suggests that all bands not directly attributable to the corkite structure arise due to atmospheric contributions, instrumental error, or a combination thereof. Standard and subsequent subtractions of this background from the acquired spectra could not be reliably performed without significant distortion of the experimental data, and thus they are presented together for comparison.

The frequencies of Raman active vibrations observed for the representative undried corkite sample agree with those reported previously on natural samples (Figure \ref{Spec_Data}b).\cite{frost2011vibrational,lafuente20151} As in the measured IR spectum, Raman bands in the synthetic sample are visibly narrower than in the naturally occurring analogue. The increased width of the vibrational bands in the natural samples is typically evidence of higher disorder in the crystal structure,\cite{Shuker1970} in agreement with the conclusion of Frost \emph{et al} that natural corkite samples deviate strongly from the ideal composition.\cite{frost2011vibrational} Band assignments are summarized in Table \ref{Raman_data} and agree well with those made by Frost \emph{et al}.\cite{frost2011vibrational} The singular discrepancy between the Raman spectrum measured from natural and synthetic samples is the band centered around 3069~cm$^{-1}$, which is not present in measurements of any jarosite-family material. Similar to the additional, lower energy absorption band present in the infrared spectrum, this feature can be understood as a consequence of the more ordered synthetic corkite structure. In natural samples, one would expect a more diffuse distribution of hydrogen bond distances throughout the structure, and thus only a single, broad band in the O-H stretching region would be expected. However, with a more ordered arrangement of alternating phosphate and sulfate groups about the kagomé sublattice, fewer distinct stretching frequencies should be represented. Application of the Libowitzky correlation, derived from a subset of bond distances in mineralogical samples, allows for an approximation of the Raman shift at which a hydroxyl stretch would be expected, given the O-O or O-H distances between the atoms participating in hydrogen bonding\cite{libowitzky1999correlation}. By this approximation, a band centered at 3069~cm$^{-1}$ would correspond to an O-O distance of 2.67 \AA, close to the 2.70~\AA~observed by Giuseppetti \emph{et al} in the \emph{R3m} structure.\cite{Giuseppetti1987corkite} This distance corresponds to the H-bonding occurring between the hydroxide doubly-coordinated by Pb and Fe and the oxygen triply-coordinated by Pb, Fe, and P. As the P-site is significantly more disordered in natural samples, hydrogen bonding of this type would occur over a more variable range of distances, and the band centered at 3069~cm$^{-1}$ would be expected to be less intense, if not absent. 

Drying of synthetic corkite samples is observed to produce significant changes in the measured Raman spectra. As shown in Figure \ref{Spec_Data}b, the intensity of the two hydroxyl vibrational modes decreases to barely above the baseline, indicative of dehydroxylation of the structure. All bridging hydroxide ligands present in undried specimens of corkite would thus become oxygen bridges upon drying. A clear broadening of the $\nu_{3}$ antisymmetric \ch{(PO4)^{3-}}/\ch{(SO4)^{2-}} modes is also observed, possibly due to changes in the coordination of the phosphorus/sulfur atom, as opposed to the more uniform environment in the less disordered, hydroxide-containing structure. Both the observed broadening of vibrational modes and higher Raman background are attributed to intrinsic photoluminescence from defects introduced as the material undergoes dehydroxylation.\cite{street1976luminescence}

Direct comparison of the measured infrared and Raman spectra for undried corkite, shown in Figure \ref{Spec_Data}c, provides additional evidence for the retention of an inversion center. By the rule of mutual exclusion, the presence of an inversion center in a structure precludes the coincident infrared and Raman activity of normal vibrational modes.\cite{venkatarayudu1954rule} If the structure lacked inversion symmetry, all normal modes would be expected to appear in both the infrared and Raman spectra, however little of this is observed for our synthetic samples. In order for pure corkite to crystallize in the non-centrosymmetric \emph{R3m} space group, the phosphate and sulfate groups were assumed to occupy discrete layers between adjacent kagomé layers. However, in both natural, non-stoichiometric samples and ordered synthetic analogues, it is much more likely that the polyanion groups instead more randomly occupy each of these layers, thereby retaining the inversion center and supporting the \emph{R$\Bar{3}$m} structural assignment.

\begin{figure}
  \includegraphics[width=1.00\textwidth]{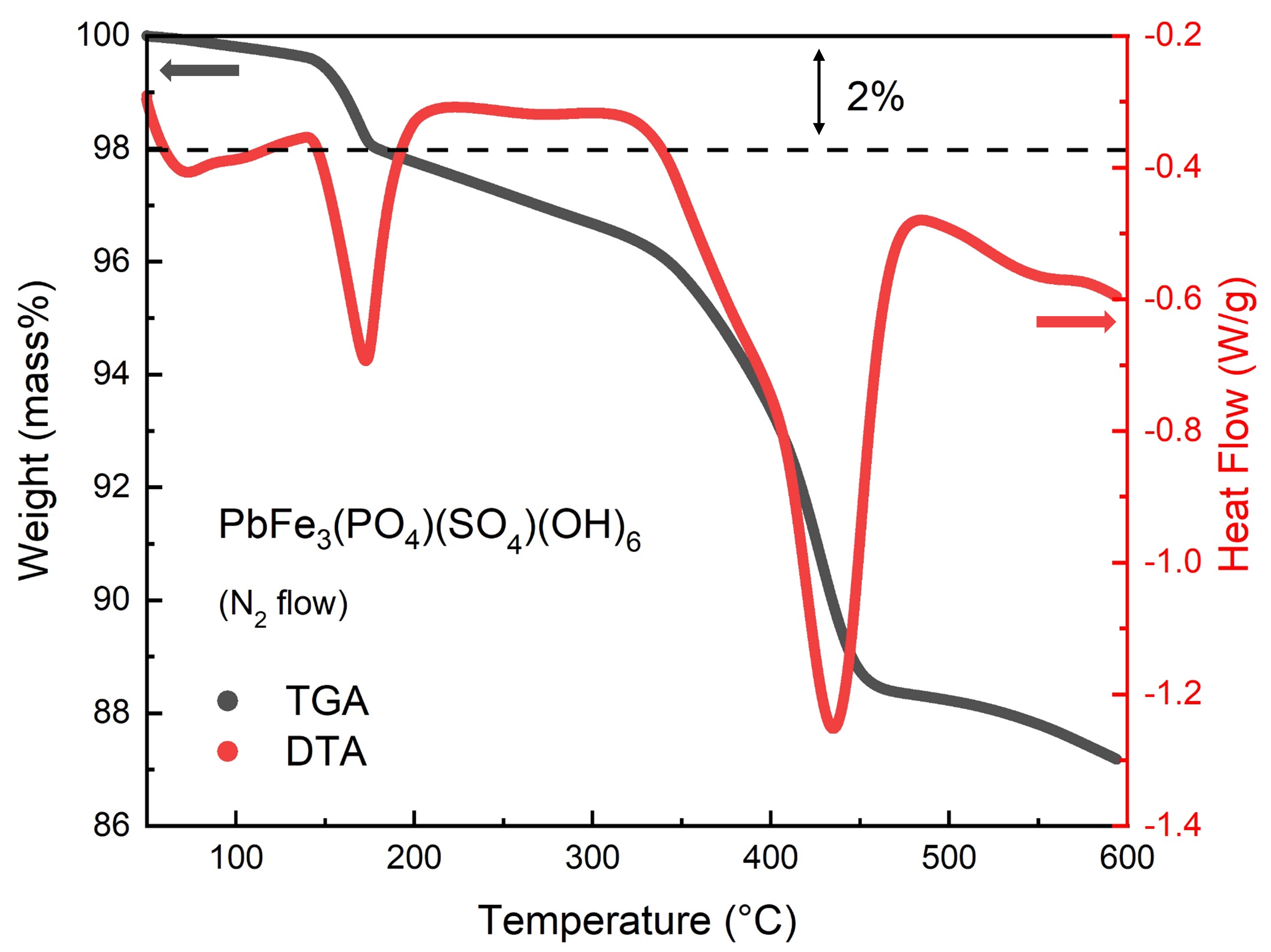}
  \caption{Thermogravimetric analysis (black) and differential thermal analysis data (red) as a function of temperature for a representative synthetic corkite powder sample. Clear signs of dehydration are present in both measurements between 140--175°C, up to 2\% of the total sample mass, or roughly 41 $\mu$mol \ch{H2O} from 58.3 $\mu$mol of total sample. The large endotherm observed above 330°C corresponds to decomposition of the corkite phase.}
  \label{TGA-DTA}
\end{figure}

TGA-DTA analysis of a representative synthetic corkite sample (Figure \ref{TGA-DTA}) was also performed to determine its stability upon heating. At around 143°C, the mass of the sample begins to decrease (up to a 2 wt\% loss), and the system is observed to absorb heat, as indicated by the endothermic peak centered around 173°C. This first set of features is most likely indicative of water being driven from the structure, as any significant decomposition would result in a much larger observed mass loss. Up to 175°C, this corresponds to a loss of approximately 41 $\mu$mol \ch{H2O} from 58.3 $\mu$mol of total sample. The second, sharper decrease in mass (totaling an additional $\sim$7.7\% loss), as well as a large endotherm, both begin above 330°C and likely correspond to decomposition of the corkite phase. This is also supported by pXRD analysis of the sample after heating to 600°C, which revealed only the presence of \ch{Fe2O3}. This trend agrees well with DTA data reported for several other synthetic jarosites by Baker, however not with that claimed to represent the behavior of synthetic corkite \cite{baker1962mineral}. Their corkite DTA measurements show no corresponding absorption of heat for the removal of water or decomposition at lower temperatures, but rather an exotherm at around 640°C, and a decomposition-type endotherm at around 950°C. This result appears rather unphysical, particularly in comparison to the typical lower thermal stability of hydrothermally-synthesized phases as well as to the data reported for other synthesized jarosites. 

\subsection{Magnetic characterization}

\begin{figure}
  \includegraphics[width=1.00\textwidth]{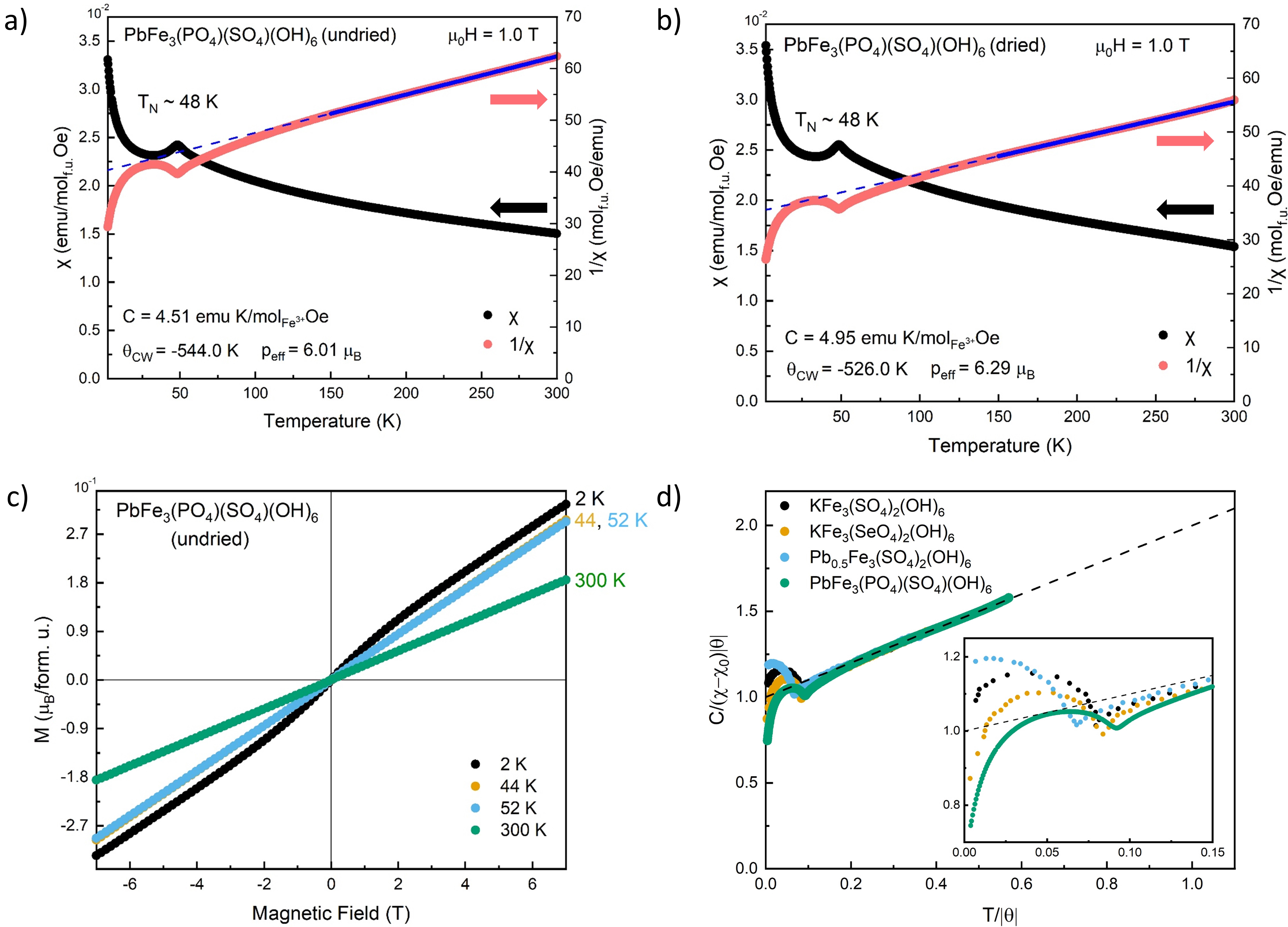}
\caption{Temperature-dependent magnetic (black) and inverse magnetic (pink) susceptibility of a representative synthetic corkite powder sample a) before and b) after drying, from 2-300 K in an applied field of 1.0 T. The best fit lines used for Curie-Weiss analysis from T~=~150-300~K are shown in blue, and extrapolated to T~=~0~K. The calculated Curie-Weiss parameters for undried [C~=~4.51(1)~emu~K/mol$_{Fe^{3+}}$~Oe, $\theta_{CW}$~=~-544.0(6)~K, p$_{eff}$~=~6.01(1)~$\mu_B$/Fe$^{3+}$, $\chi_0$~=~-0.001] and dried [C~=~4.95(1)~emu~K/mol$_{Fe^{3+}}$~Oe, $\theta_{CW}$~=~-526.0(1.1)~K, p$_{eff}$~=~6.29(1)~$\mu_B$/Fe$^{3+}$, $\chi_0$~=~-0.0025] corkite samples agree well with those reported for other Fe-deficient jarosites. c) Field-dependent magnetization of a representative undried corkite powder sample from $\mu_0$H~=~-7 to 7 T at 2 K (black), 44 K (yellow), 52 K (blue), and 300 K (green). The slight curvature in the 2 K measurement is attributed to weak, ferromagnetic intralayer coupling between adjacent \ch{Fe^{3+}} atoms. The 44 K and 52 K curves are nearly identical, and show no sign of hysteresis directly below and above the antiferromagnetic ordering temperature, T$_N$~=~48 K. Drying of the samples produced only a slight increase in the magnitude of the magnetization, and no change in the observed trend at each temperature. d) Normalized magnetization of K,S-jarosite (black)\cite{Grohol2003}, K,Se-jarosite (yellow)\cite{bartlett2005long}, Pb-jarosite (blue)\cite{bartlett2005long}, and corkite (green - this work), where C/[($\chi-\chi_0$)~$|\theta|$]~=~T/$|\theta|$~+~1. Magnetization data presented in previous works was extracted and scaled to fit the expected Curie-Weiss behavior at higher T. The inset shows the deviations of each phase from Curie-Weiss behavior from low T to their respective T$_N$; all four phases exhibit antiferromagnetic deviations from ideal behavior below T$_N$, and ferromagnetic deviations at very low T, attributable to the weak intralayer Fe-Fe coupling.}
  \label{Mag}
\end{figure}

Temperature-dependent magnetic susceptibility of a representative synthetic corkite sample shows a single, broad antiferromagnetic transition at around $T_N$~=~48 K, below those observed in the other non-hydronium Fe-jarosite phases ($T_N$~=~60-65~K) (Figure \ref{Mag}a). \cite{Wills2000} Several samples were observed to possess a second transition at around $T_N$~=~58 K, but as the lower-temperature transition was consistently present and the higher-temperature one was not, the latter was attributed to largely amorphous impurity phases not present in the measured pXRD pattern. When the as-synthesized sample mass is adjusted to account for the $\sim$~9.0\% non-magnetic \ch{Pb5(PO4)3Cl} impurity, Curie-Weiss fitting of the inverse susceptibility at high temperature (\emph{T}~=~150 K – 300 K) suggests that the iron species present is largely high-spin Fe$^{3+}$, as indicated by the calculated effective magnetic moment of p$_{eff}=$~6.01(1)~$\mu_B$. When dried overnight at 150°C, the measured magnetic susceptibility at high-temperatures (\emph{T}~$\geq$~280~K) was more linear, consistent with water being driven from the structure (Figure \ref{Mag}b). Curie-Weiss fitting for the dried sample should thus be more representative of synthetic corkite and yields an effective magnetic moment of p$_{eff}=$~6.29(1)~$\mu_B$, higher than would be expected for a high-spin Fe$^{3+}$ system, but analogous to what has been observed in other water-free synthetic jarosites \cite{bartlett2005long}. Tables comparing the magnetic properties of jarosite-family phases synthesized by both the method described in this work and by the slower, redox-based method can be found in Ref. 12 and Ref. 9, respectively. The calculated Weiss constant and frustration parameter for the dried sample, $\theta_{CW}$~=~-526.0(1.1) K and \emph{f}~=~10.96(3), indicate both strongly antiferromagnetic interactions and a high degree of frustration amongst the Fe$^{3+}$ ions on the kagomé lattice \cite{ramirez1994strongly}. 

Normalization of the magnetization as a function of temperature to the Curie-Weiss law, C/[($\chi-\chi_0$)$|\theta|$] = T/$|\theta|$~+~1, (shown in green in Figure \ref{Mag}d), exhibits ideal behavior from high temperatures to T$_N$, and antiferromagnetic deviations from T$_N$ to very low temperatures, where ferromagnetic deviations are instead observed. These deviations correspond to the onset of long-range, antiferromagnetic order and weak, intralayer ferromagnetic coupling, respectively, as has been observed in other jarosite phases. Magnetization data for synthetic \ch{KFe3(SO4)2(OH)6} (K, S-jarosite - black), \ch{KFe3(SeO4)2(OH)6} (K, Se-jarosite - yellow), and \ch{Pb_{0.5}Fe3(SO4)2(OH)6} (Pb-jarosite - blue) is also presented for comparison and agrees well with the trends observed in synthetic corkite. As in the case of corkite, the substitution of both sulfate groups in potassium jarosite for the larger selenate produces ferromagnetic deviations from the ideal Curie-Weiss behavior at low temperatures. However, these deviations are less significant in the unsubstituted \ch{KFe3(SO4)2(OH)6} data and are barely present in that of \ch{Pb_{0.5}Fe3(SO4)2(OH)6}. This  suggests that while the long-range magnetic properties of jarosite-type compounds are largely invariant upon substitution on the A-, B-, or X-site, the strength of the intralayer Fe-Fe coupling is less consistent and more heavily dependent upon the steric and electronic pressure applied on the kagomé plane by the A- and X-site cations. 

Representative magnetization measurements collected for the undried sample from $\mu_0$H = –7 T to 7 T at \emph{T}~=~2 K, 44 K, 52 K, and 300 K show no sign of magnetic hysteresis or saturation, and only a very slight curvature at \emph{T}~=~2 K from $\mu_0$H = –2 T to 2 T, supporting the absence of ferromagnetic impurities in the sample (Figure \ref{Mag}c). The slight curvature observed in the \emph{T}~=~2 K measurement can likely be attributed to weak, ferromagnetic intralayer coupling between adjacent \ch{Fe^{3+}} atoms, as has been observed in several vanadium-jarosite phases, which is fully suppressed by \emph{T}~=~35 K \cite{Papoutsakis2002}. After drying, no change was observed in the measured magnetization at any temperature, aside from a slight increase in magnitude.
The magnetic structure of pure corkite can thus likely be considered to be directly analogous to that observed across the entire family of jarosite materials, owing to the relative rigidity of the intralayer \ch{Fe^{3+}{_3}}\ch{($\mu$-OH)_3} triangular units upon substitution of the $A^{1+}$-/$A^{2+}$-, $B^{3+}$-, or $\ch{XO4}^{2-}$-/$\ch{XO4}^{3-}$-sites. Even substitution of the bounding \ch{[SO4]^{2-}} groups with higher-valent \ch{[PO4]^{3-}} produces only a 16.4° tilt of the Fe-octahedra, higher than that of the smaller A-site cation-containing jarosite phases, but slightly lower than that of the Pb-, Ag-, and Tl-jarosites. As such, the DM antisymmetric interaction which dominates the magnetic behavior of these materials is expected to produce a ferromagnetically-coupled spin-umbrella structure above each \ch{Fe^{3+}{_3}}\ch{($\mu$-OH)_3} triangle. The canted spins are then coupled antiferromagnetically between layers, producing the long-range antiferromagnetic order observed below $T_N$.

\begin{figure}
  \includegraphics[width=1.00\textwidth]{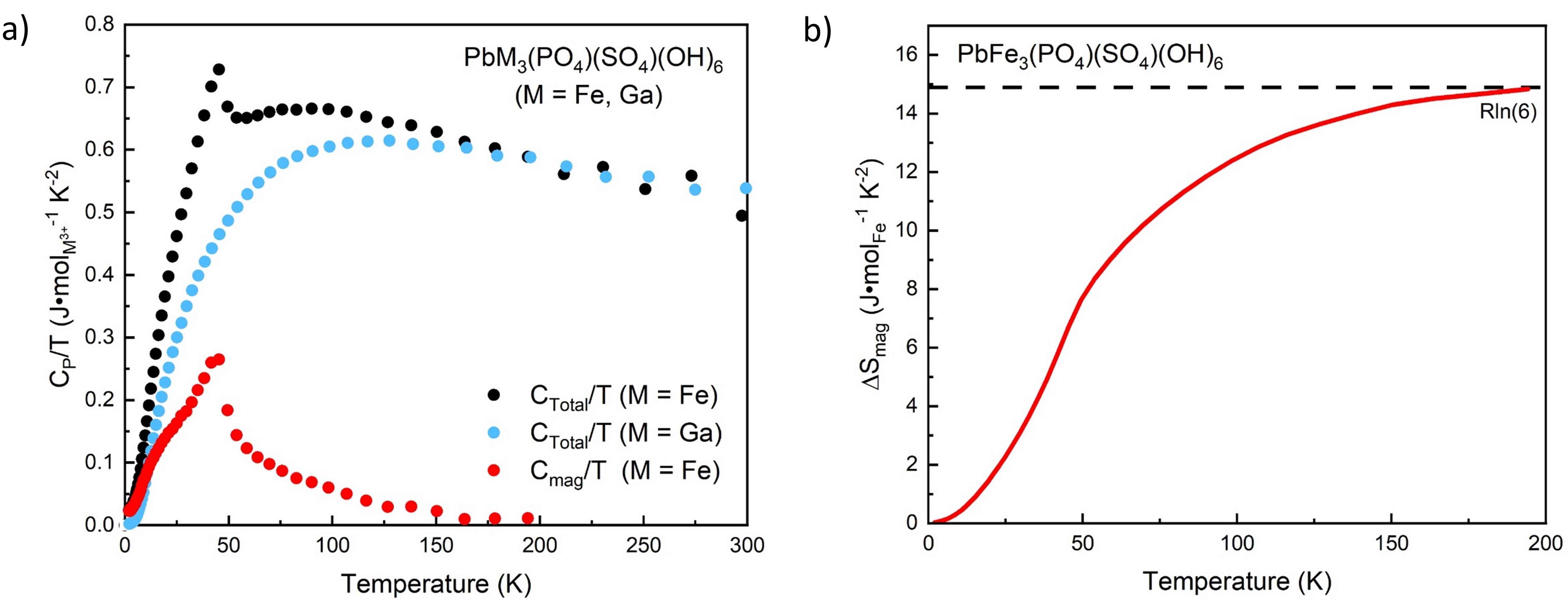}
  \caption{a) Heat capacity divided by temperature as a function of temperature, C$_{Total}$/T, for synthetic Fe-corkite (black) and Ga-corkite (blue). The measured heat capacity of Ga-corkite was scaled to account for a 4.9\% mass error. The magnetic contribution to the heat capacity of Fe-corkite, C$_{mag}$/T, as a function of temperature (red). b) Magnetic entropy rise of Fe-corkite as a function of temperature from T~=~0-200 K. The entropy rise from T~=~0~K to 2~K was estimated from linear extrapolation over this range.}
  \label{HC}
\end{figure}

Temperature-dependent heat capacity measurements of a representative synthetic Fe-corkite sample, shown divided by temperature in Figure \ref{HC}a, exhibit a peak at around \emph{T}~=~45.5 K, followed by a broader hump spanning the \emph{T}~=~50-125 K range. The measurement of less pure specimens resulted in the appearance of an additional broad feature, visible around \emph{T}~=~4-5 K, which did not correspond to any local maximum in either DC or AC susceptibility measurements. As corkite samples of representative purity did not exhibit this feature, it was attributed to the complex Pb-/Fe-phosphate impurity phase observed in some pXRD patterns. The higher-temperature fluctuations observed largely arise due to the low point density above \emph{T}~=~200 K, with the small bump at around \emph{T}~=~275 K corresponding to the freezing of the Apiezon N grease used for sample mounting.

In order to subtract the phononic contribution to the measured heat capacity, the Ga-corkite analogue was also synthesized, with a similar calculated \ch{Pb5(PO4)3Cl} impurity by mass ($\sim$9.0\%). The measured heat capacity for both samples was scaled by a factor of 0.91 to account for the phononic contribution of the impurity phase. Subtraction of the measured heat capacity of Ga-corkite from that of Fe-corkite was thus approximated as the magnetic contribution, $C_{mag}$, to the total heat capacity $C_{Total}$ of the magnetic phase. The main peak observed in both $C_{mag}$ and $C_{Total}$ agrees well with the cusp in the DC susceptibility, and is thus attributed to AFM ordering in pure corkite. Integration of $C_{mag}$ from \emph{T}~$\sim$~0-195~K yields a rise in magnetic entropy of $\Delta S_{mag}$~=~14.86~$J\cdot mol_{Fe^{3+}}^{-1}K^{-1}$ in good agreement with that expected for high-spin \ch{Fe^{3+}}, where $\Delta S_{mag}$ = Rln(2S+1) = 14.9 $J\cdot mol^{-1}K^{-1}$ for a S~=~5/2 system (Figure \ref{HC}b). Integration of the as-substracted Fe- and Ga-corkite heat capacities produced an unphysical upward trend in $\Delta S_{mag}$ up to \emph{T}~=~195 K, as opposed to the saturation expected where the thermal energy present in the system exceeds that of magnetic correlations in the structure. The inclusion of a 4.9\% mass error of the Ga-corkite sample in the calculation was found to result in the anticipated trend. Approximately 49\% of the total magnetic entropy is released below the observed ordering temperature, suggesting that short-range correlations at higher temperatures also contribute significantly to the magnetic character of stoichiometric corkite.

\section{Conclusion}
In conclusion, we report the hydrothermal synthesis and magnetic characterization of corkite, \ch{PbFe3(PO4)(SO4)(OH)6}, an understudied relative of the jarosite family of materials. Relative to previous studies on natural samples of this phase, the structure appears to have a more even and ordered distribution of phosphate and sulfate groups bounding the kagomé sublattice, as supported by the presence of additional features in and overall higher resolution of the measured infrared and Raman spectra. Magnetic susceptibility measurements of synthetic corkite reveal a transition to a long-range AFM state below $T_{N}$~=~48 K, lower than that observed in the majority of jarosite phases, indicating that the alternating bounding polyanion groups do indeed further suppress magnetic ordering in the system. The calculated effective magnetic moment and observed saturation of magnetic entropy are both on the order of what is expected for a S~=~5/2 system, indicating that Fe$^{3+}$ is the main species present on the kagomé sublattice, and that corkite's Fe-site occupancy is analogous to jarosite phases synthesized by similar methods. TGA/DTA measurements also show that the structure incorporates a small amount of water under atmospheric conditions, producing a small suppression of the calculated effective magnetic moment. Our results highlight the importance of steric strain in the development and design of novel frustrated magnetic systems, and contribute to a more complete understanding of the jarosite family of Heisenberg antiferromagnets.

\begin{acknowledgement}

This work was supported by the Institute for Quantum Matter, an Energy Frontier Research Center funded by the U.S. Department of Energy, Office of Science, Office of Basic Energy Sciences, under Grant DE-SC0019331. The TGA/DTA data was collected within the National Science Foundation Materials Innovation Platform PARADIM, under cooperative agreement \#2039380. The MPMS3 system used for magnetic characterization was funded by the National Science Foundation, Division of Materials Research, Major Research Instrumentation Program, under Grant \#1828490.

\end{acknowledgement}

\newpage

\begin{table}
    \caption{Unit-cell parameters for \ch{PbFe3(PO4)(SO4)(OH)6} at T~=~298~K, as obtained from Rietveld refinement of a pXRD pattern ($\lambda$ = 1.5406 \AA). Errors representing statistical uncertainties are shown in parentheses.}
    \label{Rietveld_table1}
    \begin{tabular}{c|c}
    \hline
       Parameter & Value \\ \hline
       Chemical Formula & \ch{PbFe3(PO4)(SO4)(OH)6} \\
       Source & Cu K$\alpha$ (1.5406 \AA) \\
       Formula weight (g/mol) & 667.8 \\
       Temperature (K) & 298 \\
       Crystal system & Rhombohedral \\
       Space group & \emph{R$\Bar{3}$m} (No. 166) \\
       a = b (\AA) & 7.3078(3) \\
       c (\AA) & 16.855(2) \\
       $\alpha$ = $\beta$ (deg) & 90 \\
       $\gamma$ (deg) & 120 \\
       Volume (\AA$^3$) & 779.53(8) \\
       Z & 1 \\
       $\rho_{calc}$ (g/cm$^3$) & 6.9243(7) \\
       R-factors (\%) & R$_P$ = 1.93, R$_{wp}$ = 2.61, R$_{exp}$ = 1.63
    \end{tabular}
\end{table}   
\begin{table}
    \caption{Atomic coordinates, occupancies, and atomic displacement parameters for \ch{PbFe3(PO4)(SO4)(OH)6} at T~=~298~K, as obtained from Rietveld refinement of a pXRD pattern ($\lambda$ = 1.5406 \AA). All occupancies besides that of iron were fixed at unity, and isotropic displacement parameters B$_{eq}$ were refined together by element type. Errors representing statistical uncertainties are shown in parentheses. Parameters listed without associated errors could not be refined without worsening of the fit to the experimental data, and thus are unchanged from previous reports.}
    \label{Rietveld_table2}
    \begin{tabular}{c|c|c|c|c|c}
    \hline
    Atom & \emph{a} (\AA) & \emph{b} (\AA) & \emph{c} (\AA) & Occupancy & $B_{eq}$ (\AA$^2$) \\\hline
    Pb & 0 & 0 & 0 & 1 & 0.54(10) \\
    Fe & 0.5013 & 0.0026 & 0.5142(7) & 0.830(6) & 0.54(10) \\
    S & 0 & 0 & -0.3070(10) & 1 & 1.8(2) \\
    P & 0 & 0 & 0.3178(11) & 1 & 1.8(2) \\
    O1 & 0 & 0 & -0.3900(21) & 1 & 0.7(2) \\ 
    O2 & 0 & 0 & 0.4099(21) & 1 & 0.7(2) \\
    O3 & -0.2191 & 0.2191 & 0.0477(18) & 1 & 0.7(2) \\
    O4 & 0.2216 & -0.2216 & -0.0581(19) & 1 & 0.7(2) \\
    O5 & -0.1264 & 0.1264 & -0.1206(14) & 1 & 0.7(2) \\
    O6 & 0.1264 & -0.1264 & 0.1374(16) & 1 & 0.7(2) \\
    H1 & -0.0162(867) & -0.0176(671) & -0.0018(1073) & 1 & 1 \\
    H2 & 0.3333(150) & -0.2097(72) & 0.1551(125) & 1 & 1 \\\hline
    \end{tabular}
\end{table}

\begin{table}
    \caption{Assignment of observed infrared stretching frequencies}
    \label{IR_data}
    \centering
    \begin{tabular}{c|c|c}
    \hline
       Band center(s) (cm$^{-1}$) & Band strength & Vibrational mode(s) \\ \hline
       591, 624 & Weak & $\nu_{4}$ \ch{(SO4)^{2-}} bending modes\\ 
       725 & Broad & $\delta$-\ch{FeOH} deformation modes \\
       955 & Strong & \ch{(PO4)^{3-}} symmetric stretch \\
       1060, 1174 & Strong, Medium & \ch{(PO4)^{3-}}/\ch{(SO4)^{2-}} antisymmetric stretch \\ 
       1637 & Weak & Water \\
       3315 & Broad & \ch{OH-} stretch \\
    \end{tabular}
\end{table}

\begin{table}
    \caption{Assignment of observed Raman stretching frequencies}
    \label{Raman_data}
    \centering
    \begin{tabular}{c|c|c}
    \hline
       Band center(s) (cm$^{-1}$) & Band strength & Vibrational mode(s) \\ \hline
       110, 145, 229 & Medium, Strong & FeO stretch \\ 
       346 & Medium & $\nu_{2}$ \ch{(PO4)^{3-}} \\
       437 & Strong & $\nu_{2}$ \ch{(SO4)^{2-}} \\
       555-571 & Medium & $\nu_{4}$ \ch{(PO4)^{3-}} \\
       622 & Medium & $\nu_{4}$ \ch{(SO4)^{2-}} \\
       1005 & Strong & $\nu_{1}$ \ch{(SO4)^{2-}} \\
       1103-1162 & Medium & $\nu_{3}$ \ch{(PO4)^{3-}}/$\nu_{3}$ \ch{(SO4)^{2-}} \\
       3069, 3390 & Medium, Broad & $\nu$ \ch{OH-} \\
    \end{tabular}
\end{table}

\newpage

\newpage

\clearpage
\bibliography{Refs.bib}

\end{document}